\documentclass[twocolumn]{aastex61}

\submitjournal{ApJL}

\shorttitle{Neutron-star radius constraints}
\shortauthors{Bauswein et al.}
\begin{document}

\title{Neutron-star radius constraints from GW170817 and future detections}
 
\correspondingauthor{Andreas Bauswein}
\email{andreas.bauswein@h-its.org}

\author{Andreas Bauswein}
\affil{Heidelberger Institut f\"ur Theoretische Studien, Schloss-Wolfsbrunnenweg 35, D-69118 Heidelberg, Germany}

\author{Oliver Just}
\affil{Astrophysical Big Bang Laboratory, RIKEN, Saitama 351-0198, Japan}

\author{Hans-Thomas Janka}
\affil{Max-Planck-Institut f\"ur Astrophysik, Karl-Schwarzschild-Str. 1, D-85748 Garching, Germany}

\author{Nikolaos Stergioulas}
\affil{Department of Physics, Aristotle University of Thessaloniki, GR-54124 Thessaloniki, Greece}

\begin{abstract}
We introduce a new, powerful method to constrain properties of neutron stars (NSs). We show that the total mass of GW170817 provides a reliable constraint on the stellar radius if the merger did not result in a prompt collapse as suggested by the interpretation of associated electromagnetic emission. The radius $R_{1.6}$ of nonrotating NSs with a mass of 1.6~$M_\odot$ can be constrained to be larger than 
$10.68_{-0.04}^{+0.15}$~km, and the radius $R_\mathrm{max}$ of the nonrotating maximum mass configuration must be larger than
$9.60_{-0.03}^{+0.14}$~km. We point out that detections of future events will further improve these constraints. Moreover, we show that a future event with a signature of a prompt collapse of the merger remnant will establish even stronger constraints on the NS radius from above and the maximum mass $M_\mathrm{max}$ of NSs from above. These constraints are particularly robust because they only require a measurement of the chirp mass and a distinction between prompt and delayed collapse of the merger remnant, which may be inferred from the electromagnetic signal or even from the presence/absence of a ringdown gravitational-wave (GW) signal. %As more events are observed the distinction may become easier from the interpretation of the larger sample of electromagnetic counterparts and from more advanced theoretical models of the emission. 
This prospect strengthens the case of our novel method of constraining NS properties, which is directly applicable to future GW events with accompanying electromagnetic counterpart observations. We emphasize that this procedure is a new way of constraining NS radii from GW detections independent of existing efforts to infer radius information from the late inspiral phase or postmerger oscillations, and it does not require particularly loud GW events.
\end{abstract}

%% Keywords should appear after the \end{abstract} command. 
%% See the online documentation for the full list of available subject
%% keywords and the rules for their use.

%% From the front matter, we move on to the body of the paper.
%% Sections are demarcated by \section and \subsection, respectively.
%% Observe the use of the LaTeX \label
%% command after the \subsection to give a symbolic KEY to the
%% subsection for cross-referencing in a \ref command.
%% You can use LaTeX's \ref and \label commands to keep track of
%% cross-references to sections, equations, tables, and figures.
%% That way, if you change the order of any elements, LaTeX will
%% automatically renumber them.

%% We recommend that authors also use the natbib \citep
%% and \citet commands to identify citations.  The citations are
%% tied to the reference list via symbolic KEYs. The KEY corresponds
%% to the KEY in the \bibitem in the reference list below. 

% Abkuerzungen:
% BH
% NS
% GW
% EoS
% GRB
% SNR
%

\section{Introduction} \label{sec:intro}
The recently detected GW170817 is the first observed gravitational-wave (GW) source involving matter \citep{Abbott2017}. The measured binary masses point to a merger of two neutron stars (NSs). %The measured binary masses are only compatible with a neutron-star (NS) merger. 
Apart from the importance of this detection for stellar astrophysics and nucleosynthesis, such events are highly interesting because they bear the potential to infer weakly-constrained properties of NSs \citep{Lattimer2016,Oezel2016,Oertel2017}. Such information can be obtained from the GW signal either from finite-size effects during the late inspiral phase \citep[e.g.][]{Faber2002,Flanagan2008,Read2013,DelPozzo2013,Abbott2017} or through the characteristic oscillations of the postmerger remnant \citep{Bauswein2012,Bauswein2012a,Bauswein2014,Takami2014,Clark2014,Chatziioannou2017}. Both approaches require high signal-to-noise ratios (SNRs).% The relative proximity of GW170817 and the therefore presumably high NS merger rate suggest that such measurements might be possible already in the era of the current GW detectors.

The merging of two NSs can result either in the direct formation of a black hole (BH) on a dynamical time scale (prompt collapse) or the formation of an at least transiently stable NS merger remnant (delayed/no collapse). The former case occurs for mergers with binary masses $M_\mathrm{tot}$ above a threshold binary mass $M_\mathrm{thres}$, a delayed or no collapse results for binaries with $M_\mathrm{tot}<M_\mathrm{thres}$. The two different collapse scenarios are also expected to lead to different electromagnetic emission. The amount of dynamical ejecta is strongly reduced in the case of prompt BH formation \citep{Bauswein2013a,Hotokezaka2013}. Also, the different nature of the merger remnant yields different amounts of secular ejecta \citep{Fernandez2013,Metzger2014,Perego2014,Siegel2014,Just2015}.

In this letter we present a new method to infer information on the NS equation of state (EoS) from NS mergers that does not require a high SNR of the GW measurement. Our constraint only relies on the measured binary mass of GW170817 and the evidence for a delayed/no collapse in this event as suggested by its electromagnetic emission \citep[e.g.][]{Kasen2017,Metzger2017}.
In the case of a delayed/no collapse the measured total binary mass of GW170817 provides a lower bound on the threshold mass for direct BH formation,
\begin{equation}\label{eq:limit}
M_\mathrm{thres} > M_\mathrm{tot}^\mathrm{GW170817}= 2.74_{-0.01}^{+0.04}~M_\odot,
\end{equation}
and we conclude that the radius $R_{1.6}$ of a NS with 1.6~$M_\odot$ must be larger than 
$10.68_{-0.04}^{+0.15}$~km. We demonstrate that our new method promises very strong constraints on NS radii and the maximum mass $M_\mathrm{max}$ of nonrotating NSs if more NS mergers will be observed and in particular if an event with a prompt collapse of the merger remnant is identified. %[Our new procedure is directly applicable to any new GW event which allows a distinction between prompt and delayed/no collapse e.g. by accompanying electromagnetic emission, while it does not require a high SNR of the GW signal. The method has already been partially anticipated in \cite{Bauswein2013,Bauswein2016}]

\begin{figure*}[t]

\includegraphics[width=9cm]{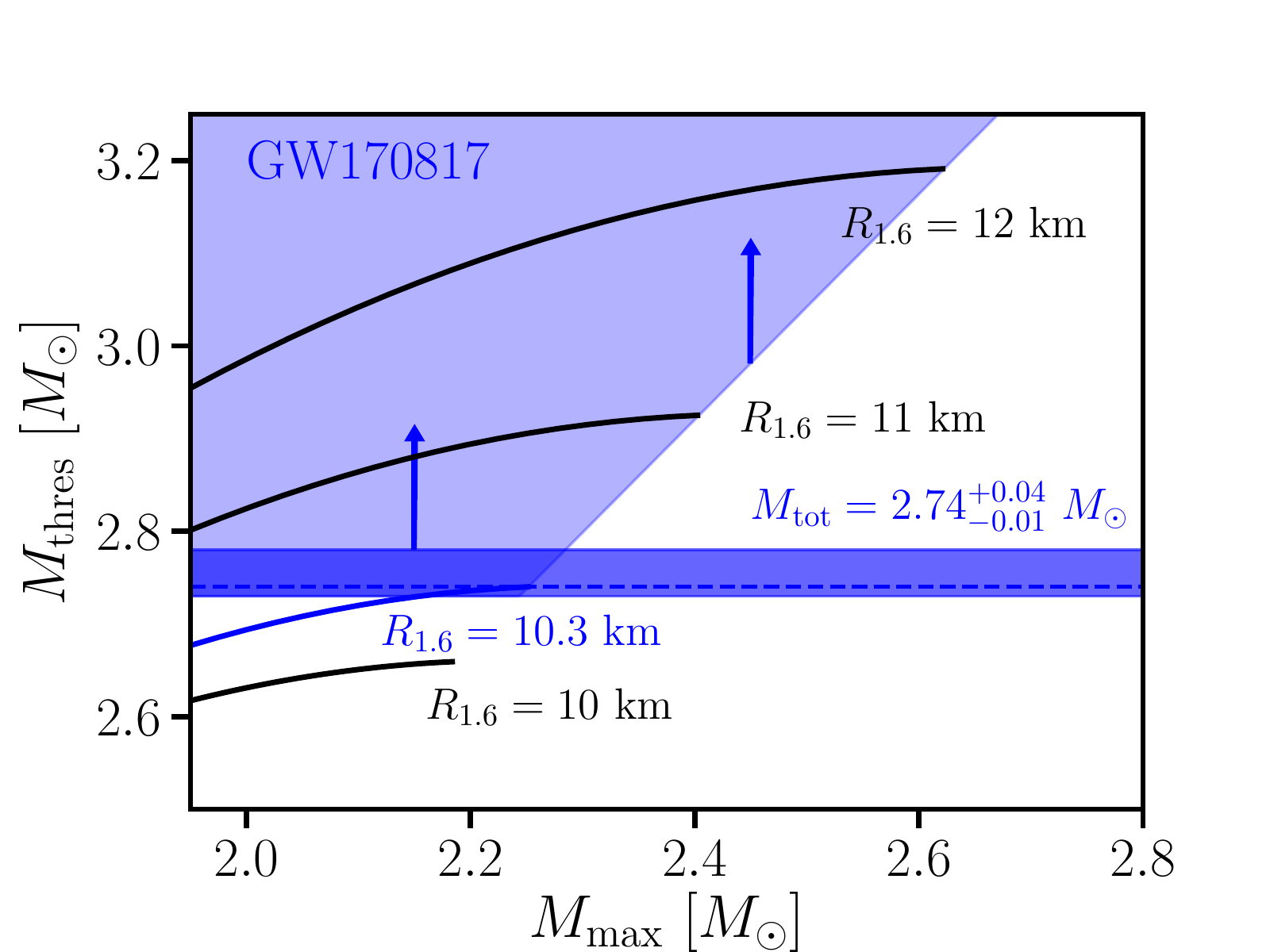}\includegraphics[width=9cm]{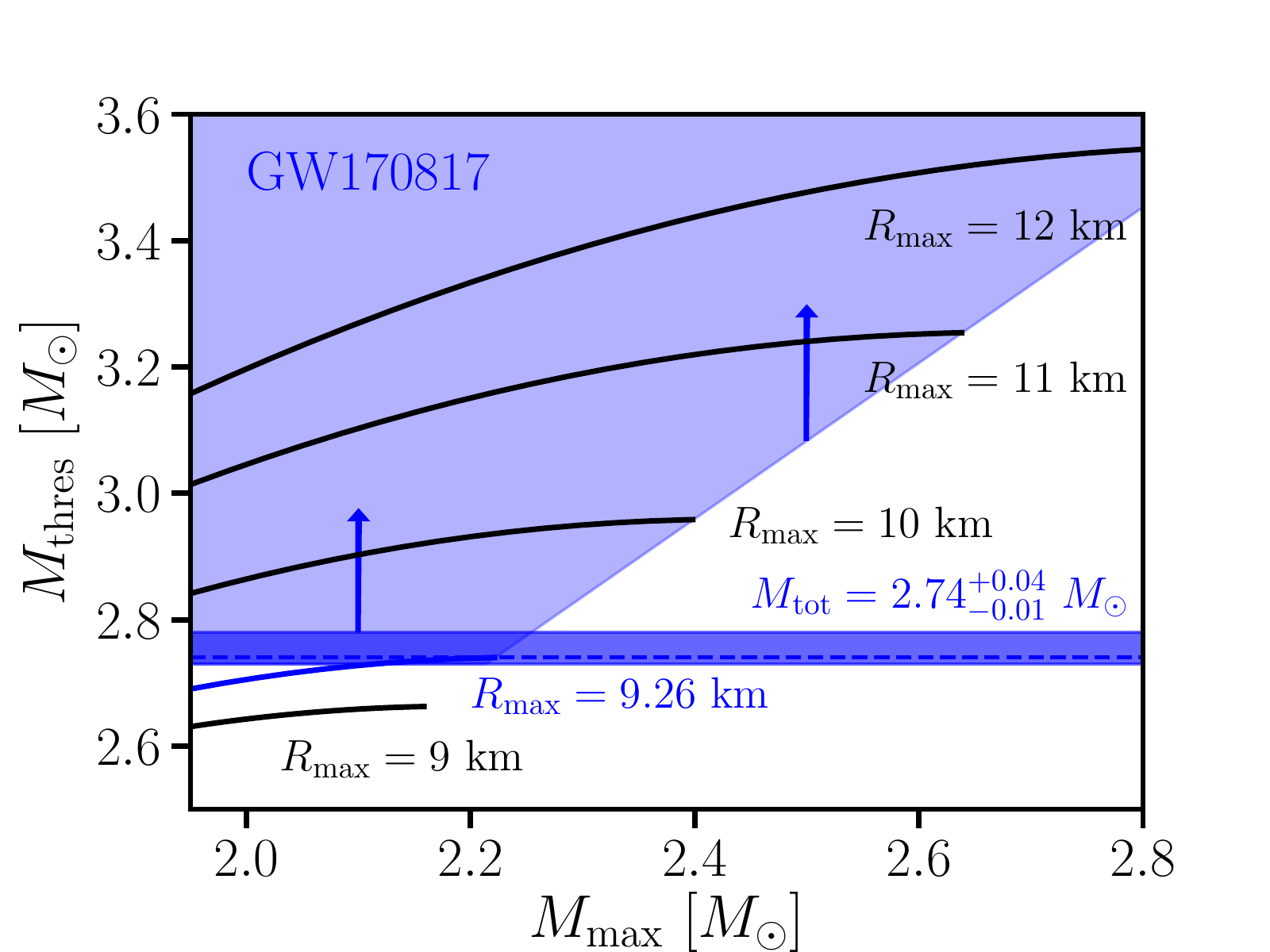}
\caption{Threshold binary mass $M_\mathrm{thres}$ for prompt collapse as function of $M_\mathrm{max}$ for different $R_{1.6}$ (left panel, Eq.~\ref{eq:mthrr16}) and $R_\mathrm{max}$ (right panel, Eq.~\ref{eq:mthrrmax}) (solid lines). The dark blue band shows the total binary mass of GW170817 providing a lower limit on $M_\mathrm{thres}$. The true $M_\mathrm{thres}$ must lie within the light blue areas if GW170817 resulted in a delayed/no collapse. This rules out NSs with $R_{1.6}\leq 10.30_{-0.03}^{+0.18}$~km and $R_\mathrm{max} \leq 9.26_{-0.03}^{+0.17}$~km. Causality requires $M_\mathrm{thres}\geq 1.22 M_\mathrm{max}$ (left panel) and $M_\mathrm{thres}\geq 1.23 M_\mathrm{max}$ (right panel).\label{fig:mthresmmax}}

\end{figure*}

\section{Observations}
Several telescopes observed emission in the X-ray, optical and infrared from the GW source with spatial and temporal coincidence \citep{Multi2017}. The observations are compatible with NS merger ejecta that are heated by the nuclear decays of products of the rapid neutron-capture process \citep{Metzger2010}. The light-curve properties were interpreted as being produced by dynamical ejecta from the merger and secular ejecta from the merger remnant. The estimated ejecta mass is in the range 0.03 to 0.05~$M_\odot$ \citep{Cowperthwaite2017,Kasen2017,Nicholl2017,Chornock2017,Drout2017,Smartt2017,Kasliwal2017,Kilpatrick2017,Tanvir2017,Tanaka2017}, which even for asymmetric binaries lies near the high end of the theoretical range expected from simulations. %Bla et al. distinguished the emission from dynamical red and secular blue ejecta and inferred corresponding masses in the range of 0.03-0.04~$M_\odot$ and 0.01-0.025~$M_\odot$ respectively. 
This can be interpreted as tentative evidence for a delayed/no collapse in GW170817 because this merger outcome tends to produce larger ejecta masses as compared to a direct collapse \citep{Bauswein2013a,Hotokezaka2013,Fernandez2013,Metzger2014,Perego2014,Siegel2014,Wanajo2014,Just2015,Sekiguchi2016}. We thus use below the assumption of no prompt collapse in GW170817 and leave the detailed interpretation of the electromagnetic emission to future work. Our assumption can be corroborated by refined models and future observations.

%Similar arguments have been presented e.g. in \cite{Metzger2017,Margalit2017}.

\section{Neutron-star radius constraints}\label{sec:radconst}
\subsection{Threshold binary mass}
If GW170817 resulted in a delayed collapse or no collapse, its total mass provides a lower limit on the threshold binary mass for prompt collapse as given by Eq.~(\ref{eq:limit}).%$M_\mathrm{thres}$ that distinguishes direct and delayed/no BH formation. 
% \begin{equation}\label{eq:limit}
% M_\mathrm{thres} > M_\mathrm{tot}=2.74_{-0.01}^{+0.04}~M_\odot.
% \end{equation}

The threshold binary mass $M_\mathrm{thres}$ depends sensitively on the EoS \citep{Shibata2005,Baiotti2008,Hotokezaka2011,Bauswein2013}. Considering different EoSs, in~\cite{Bauswein2013} we found by hydrodynamical simulations that the threshold binary mass to good accuracy follows
\begin{equation}\label{eq:mthrr16}
M_\mathrm{thres}=\left( -3.606\frac{G M_\mathrm{max}}{c^2 R_{1.6}} +2.38 \right) \cdot M_\mathrm{max}
\end{equation}
with $R_{1.6}$ being the radius of a nonrotating NS with a mass of 1.6~$M_\odot$ and $M_\mathrm{max}$ being the maximum mass of nonrotating NSs. The relation was derived from simulations of symmetric binary mergers but also holds for moderately asymmetric systems \citep{Bauswein2013,Bauswein2017}. We verify by additional simulations for 5 representative EoSs that strongly asymmetric mergers with mass ratio $q=0.6$ have a threshold binary mass which is systematically lower by 0.1 to 0.3~$M_\odot$ than $M_\mathrm{thres}$ of equal-mass binaries. This reduction of $M_\mathrm{thres}$ for asymmetric binaries is understandable  because according to Kepler's law asymmetric binaries have less angular momentum than equal-mass binaries with the same $M_\mathrm{tot}$ at a given orbital separation, which implies less stabilization for asymmetric mergers. (With the low-spin priors the 90\% credibility interval of the mass ratio of GW170817 is $q=0.7-1.0$). % The heavier binary component forming the core of the merger remnant moves more slowly on its orbit and thus the specific angular momentum in the core is relatively low, which results in less stabilization. 
If GW170817 was very asymmetric, one has $M^\mathrm{asym}_\mathrm{thres}\geq M_\mathrm{tot}$, which implies that Eq.~(\ref{eq:limit}) is conservative because $M_\mathrm{thres}>M^\mathrm{asym}_\mathrm{thres}$.% for a given $R_{1.6}$. Avoiding a prompt collapse in the asymmetric case would require an even larger value of $R_{1.6}$ than for symmetric mergers.

%Eq.~(\ref{eq:mthrr16}) is accurate to better than $0.1~M_\odot$ (see \cite{Bauswein2013,Bauswein2016} for details).

A similarly accurate description of $M_\mathrm{thres}$ is given by the fit
\begin{equation}\label{eq:mthrrmax}
M_\mathrm{thres}=\left( -3.38\frac{G M_\mathrm{max}}{c^2 R_\mathrm{max}} +2.43 \right) \cdot M_\mathrm{max}
\end{equation}
with the radius $R_\mathrm{max}$ of the maximum-mass configuration. Eq.~(\ref{eq:mthrr16}) is accurate to better than $0.1~M_\odot$ \citep{Bauswein2013,Bauswein2016}. The existence of these relations has been solidified by semi-analytic calculations of equilibrium models \citep{Bauswein2017}.%[These relations show that $M_\mathrm{thres}$ does not only grow with $M_\mathrm{max}$, but that the threshold also depends on NS radii. For EoSs with the same $M_\mathrm{max}$, the EoSs with larger radii result in higher $M_\mathrm{thres}$.]
%\newpage
\subsection{Radius constraints}
Equations~(\ref{eq:mthrr16}) and~(\ref{eq:mthrrmax}) imply constraints on NS radii $R_{1.6}$ and $R_\mathrm{max}$ since the total binary mass of GW170817 represents a lower bound on $M_\mathrm{thres}$ (Eq.~(\ref{eq:limit})). Figure~\ref{fig:mthresmmax} (left panel) shows $M_\mathrm{thres}(M_\mathrm{max}; R_{1.6})$ (Eq.~(\ref{eq:mthrr16})) for different chosen values of $R_{1.6}$ (solid lines). Every sequence terminates at
\begin{equation}
M_\mathrm{max}=\frac{1}{3.10}\,\frac{c^2 R_{1.6}}{G},
\end{equation}
which is an empirical %safe 
upper limit on $M_\mathrm{max}$ for the given $R_{1.6}$. Extending various microphysical EoSs with a maximally stiff EoS, i.e. $v_\mathrm{sound}=c$, beyond the central density of a NS with $1.6~M_\odot$ determines the highest possible $M_\mathrm{max}$ for a given $R_{1.6}$ compatible with causality. With Eq.~(\ref{eq:mthrr16}) it implies $M_\mathrm{thres}\geq 1.22 M_\mathrm{max}$.

In Fig. \ref{fig:mthresmmax} the horizonal dark blue band refers to the measured lower limit of $M_\mathrm{thres}$ given by the total binary mass of GW170817 (Eq.~(\ref{eq:limit})). This GW measurement thus rules out EoSs with very small $R_{1.6}$ because those EoSs would not result in a delayed collapse for the measured binary mass. The allowed range of possible stellar parameters is indicated by the light blue area. The solid blue curve corresponds to the smallest $R_{1.6}$ compatible with Eq.~(\ref{eq:limit}). Hence, the radius of a 1.6~$M_\odot$ NS must be larger than $10.30_{-0.03}^{+0.15}$~km. The error bar corresponds to the radii compatible with the error in $M_\mathrm{tot}$. Arguments about the error budget and the robustness are provided in Sect.~\ref{sec:discussion}. 

%The relation in Eq.~(\ref{eq:mthrrmax}) can be employed to obtain a limit on $R_\mathrm{max}$. 
Figure~\ref{fig:mthresmmax} (right panel) displays $M_\mathrm{thres}(M_\mathrm{max}; R_\mathrm{max})$ for different chosen $R_\mathrm{max}$ (solid lines). The different sequences for fixed $R_\mathrm{max}$ are constrained by causality \citep{Koranda1997,Lattimer2016} requiring
\begin{equation}\label{eq:causal}
M_\mathrm{max} \leq \frac{1}{2.82}\frac{c^2 R_\mathrm{max}}{G}
\end{equation}
and with Eq.~(\ref{eq:mthrrmax})
\begin{equation}
M_\mathrm{thres} \geq 1.23\, M_\mathrm{max}.
\end{equation}

The lower bound of $M_\mathrm{thres}$ given by the measured total mass of GW170817 is shown as dark blue band. The radius $R_\mathrm{max}$ of the nonrotating maximum-mass NS is thus constrained to be larger than $9.26_{-0.03}^{+0.17}$~km.

Instead of using Eq.~(\ref{eq:limit}) it may be more realistic to assume that the remnant was stable for at least 10~milliseconds to yield the observed ejecta properties (high masses, blue component) \citep{Margalit2017,Nicholl2017,Cowperthwaite2017}. In this case our numerical simulations suggest that $M_\mathrm{thres}-M_\mathrm{tot}\geq 0.1~M_\odot$. This strengthens the radius constraints to $R_{1.6}\geq 10.68_{-0.04}^{+0.15}$~km and $R_\mathrm{max}\geq 9.60_{-0.03}^{+0.14}$~km.

\begin{figure}[ht!]  %tovplotafs.py
\includegraphics[width=9.7cm]{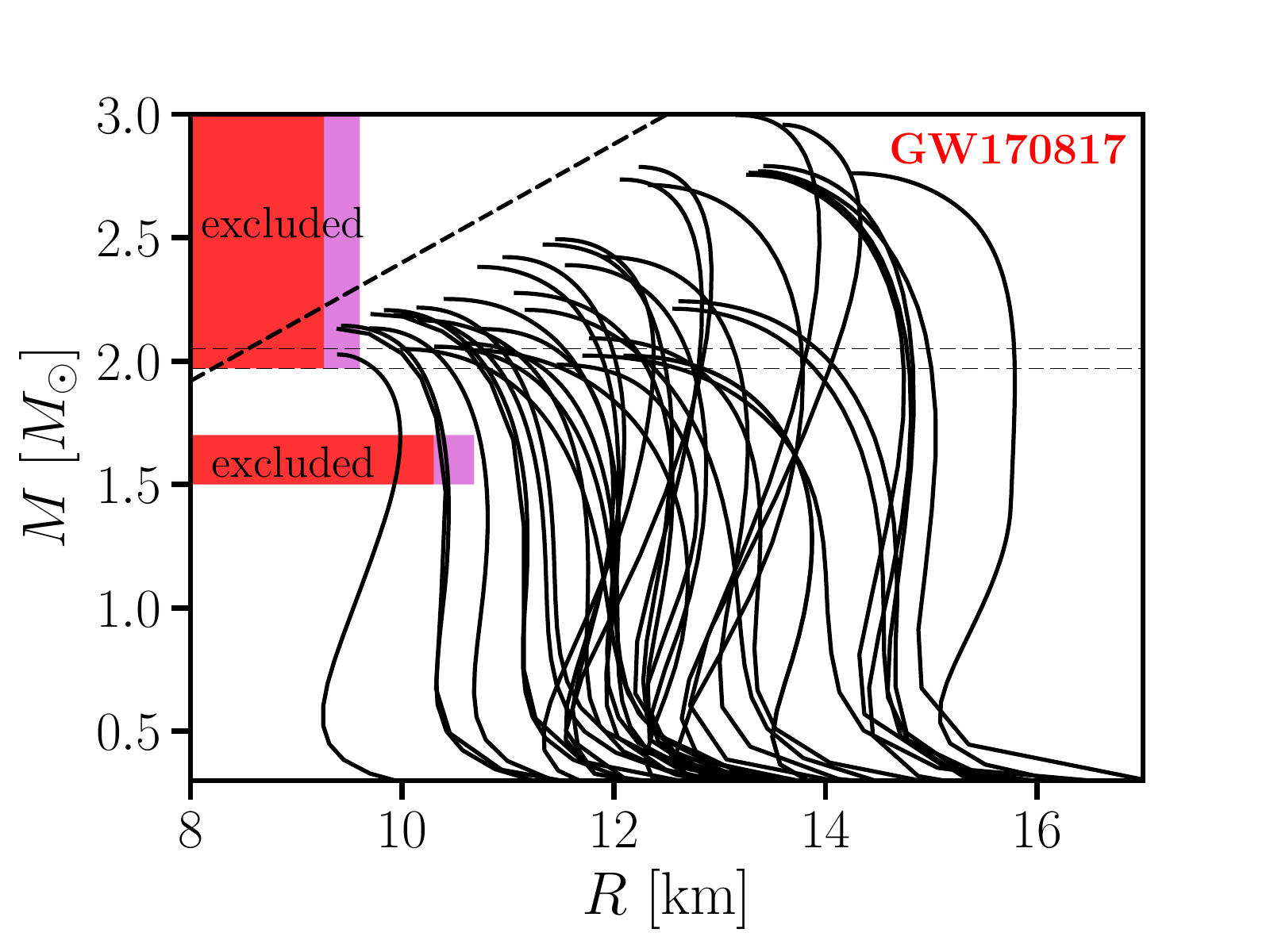}
\caption{Mass-radius relations of different EoSs with very conservative (red area) and ``realistic'' (cyan area) constraints of this work for $R_{1.6}$ and $R_\mathrm{max}$. Horizontal lines display the limit by \cite{Antoniadis2013}. The dashed line shows the causality limit.\label{fig:tov}}
\end{figure}

Figure~\ref{fig:tov} shows these radius constraints overlaid on mass-radius relations of different EoSs available in the literature. 
Our new radius constraints for $R_{1.6}$ and $R_\mathrm{max}$ derived from GW170817 exclude EoS models describing very soft nuclear matter. For the three EoSs excluded by our ``realistic'' constraint in Fig.~\ref{fig:tov}, e.g. the softest EoS in~\cite{Hebeler2013}, we crosschecked that numerical simulations with the binary masses of GW170817 do indeed result in a prompt collapse.

%\newcolumn
%\columnbreak
\subsection{Discussion: robustness and errors}\label{sec:discussion}
We took an overall conservative approach in this first study. Future refinements may strengthen these constraints. Our way of inferring NS radii is particularly appealing and robust because it only relies on (1) a well measured quantity (total binary mass with reliable error bars), (2) a single verifiable empirical relation (Eqs.~(\ref{eq:mthrr16}) or~(\ref{eq:mthrrmax})) derived from simulations, and (3) a clearly defined working hypothesis (delayed/no collapse of the merger remnant). All assumptions can be further substantiated and refined by more advanced models and future observations, and error bars can be robustly quantified.

(1) Mass measurement: The total binary mass can be measured with good accuracy and the error bars are given with high confidence. We fully propagate the error through our analysis using the low-spin prior results of \cite{Abbott2017}. If GW170817 was an asymmetric merger as tentatively suggested by the high ejecta mass, the true $M_\mathrm{tot}$ lies at the upper bound of the error band and our radius constraints become stronger.

(2) Accuracy of empirical relations for $M_\mathrm{thres}$: The empirical relations (Eqs.~(\ref{eq:mthrr16}) and~(\ref{eq:mthrrmax})) are inferred from hydrodynamical simulations \citep{Bauswein2013,Bauswein2016} and carry a systematic error\footnote{Simulations for determining $M_\mathrm{thres}$ and corresponding fits employ a conformally flat spatial metric with a GW backreaction scheme \citep{Oechslin2007,Bauswein2013}, which results in a slightly decelerated inspiral (compared to fully relativistic calculations) and thus leads to a slight overestimation of $M_\mathrm{thres}$ by $\sim 0.05~M_\odot$. We will quantify this effect in future work and emphasize that a small overestimation implies that our radius constraints are conservative.} and an intrinsic scatter (stemming from the sample of candidate EoSs, which do not perfectly fulfill the analytic fit). $M_\mathrm{thres}$ has been numerically determined with a precision of $\pm0.05~M_\odot$. Deviations between fits and numerical data are on average less than $0.03~M_\odot$ and at most $0.075~M_\odot$\footnote{We computed $M_\mathrm{thres}$ for six additional EoSs not included in~\cite{Bauswein2013} to verify this accuracy in particular for EoS models yielding relatively small NS radii (as small as $R_{1.6}=10.37$~km).}. We do not include this uncertainty in our error analysis because the numerically determined $M_\mathrm{thres}$ of all tested microphysical candidate EoSs is significantly smaller than the maximum of the $M_\mathrm{thres}(M_\mathrm{max})$ sequence for the radius given by the respective EoS\footnote{Within our sample of 17 candidate EoSs the true $M_\mathrm{thres}$ is on average $0.17~M_\odot$ ($0.14~M_\odot$ for the $R_\mathrm{max}$ sequence) below the maximum $M_\mathrm{thres}^\mathrm{up}$ of the $M_\mathrm{thres}(M_\mathrm{max},R)$ relation, which well justifies to neglect the scatter in Eqs.~(\ref{eq:mthrr16}) and~(\ref{eq:mthrrmax}). Three EoSs (eosAU, WFF1, LS375) are relatively close to the maximum ($\sim 0.02~M_\odot$ below $M_\mathrm{thres}^\mathrm{up}$). However, these EoS models become acausal ($v_\mathrm{sound}>c$), i.e. unrealistically stiff, at densities of high-mass merger remnants, which artificially increases $M_\mathrm{thres}$. For these EoSs we determined $M_\mathrm{thres}$ with a precision of $\pm 0.025~M_\odot$.}. Recall that the maxima of the $M_\mathrm{thres}(M_\mathrm{max})$ sequences are given by maximally (unrealistically) stiff EoSs only constrained by causality. We thus remain conservative by determining minimum NS radii through the maxima of the sequences defined by causality.

We note that evidence for a long-lived merger remnant \citep[e.g.][]{Lippuner2017,Margalit2017} further strengthens our arguments. The longer the remnant lifetime $\tau$, the larger is the difference $M_\mathrm{thres}-M_\mathrm{tot}>0$, which implies stronger radius constraints (see above). These considerations emphasize the importance of a better understanding of the dependence of the remnant lifetime on the binary mass, which represents a challenge for numerical simulations, but could yield even stronger radius constraints (see Sect.~\ref{sec:future}). Currently, the lifetime of presumably more than just a few milliseconds for the remnant in GW170817 implies an additional buffer in our error analysis.

The validity of Eqs.~(\ref{eq:mthrr16}) and~(\ref{eq:mthrrmax}) and their uncertainties should be explored by future simulations employing an even larger set of EoSs (including models of absolutely stable quark matter) and successively improved numerical modeling. %Since the numerical data for $M_\mathrm{thres}$ are also compatible with different analytic expressions for the fit, we verified that the exact form of the fit does not significantly affect our results. %
%It should be checked whether the empirical relations hold for absolutely stable strange stars and scenarios involving different families of compact stars. %The existence of relations~(\ref{eq:mthrr16}) and~(\ref{eq:mthrrmax}) has been solidified by semi-analytic calculations of equilibrium models \citep{Bauswein2017}. %[Overall, our procedure for radius constraints introduced in this work has been designed to be conservative to arrive at robust constraints, while leaving room for further refinements.]

Obviously, the merger outcome for a given EoS can be directly tested through numerical simulations for the measured binary masses to validate our constraints.

(3) Distinction of collapse scenarios: The scenario of a delayed/no collapse in GW170817 can be consolidated by more advanced models of the electromagnetic emission. %involving hydrodynamical merger simulations, nuclear network calculations and radiative transfer calculations. Moreover, w
We anticipate that as more GW and counterpart observations become available in the future, the comprehension of their emission features will grow and will allow a more robust distinction between prompt and delayed collapse events. %[For instance, a steep drop in the luminosity of different events with increasing binary mass could point to the threshold between prompt and delayed collapse (see, e.g. Figs. 7 and 11 in \cite{Bauswein2013a}).] 
The growing understanding can be applied to the interpretation of past events by using additional information about the remnant lifetime for continuous refinements of the radius constraints. The interpretation of electromagnetic emission resulting from prompt or delayed collapse can be tested in the future also by measuring postmerger GW emission \citep{Clark2014}.

\section{Future measurements}\label{sec:future}
Ideas introduced in this paper bear the potential of very strong EoS constraints as they are applied to future GW events with higher binary masses. %[In particular, these detections do not require a higher SNR for further constraints, which emphasizes the power of our method.] It is natural to expect that, as more observations become availabe, the distinction between prompt and delayed/no collapse events will become more robust as the sample of observed counterparts grows. 
We point out three future hypothetical scenarios.

\begin{figure}[ht!]
\includegraphics[width=9cm]{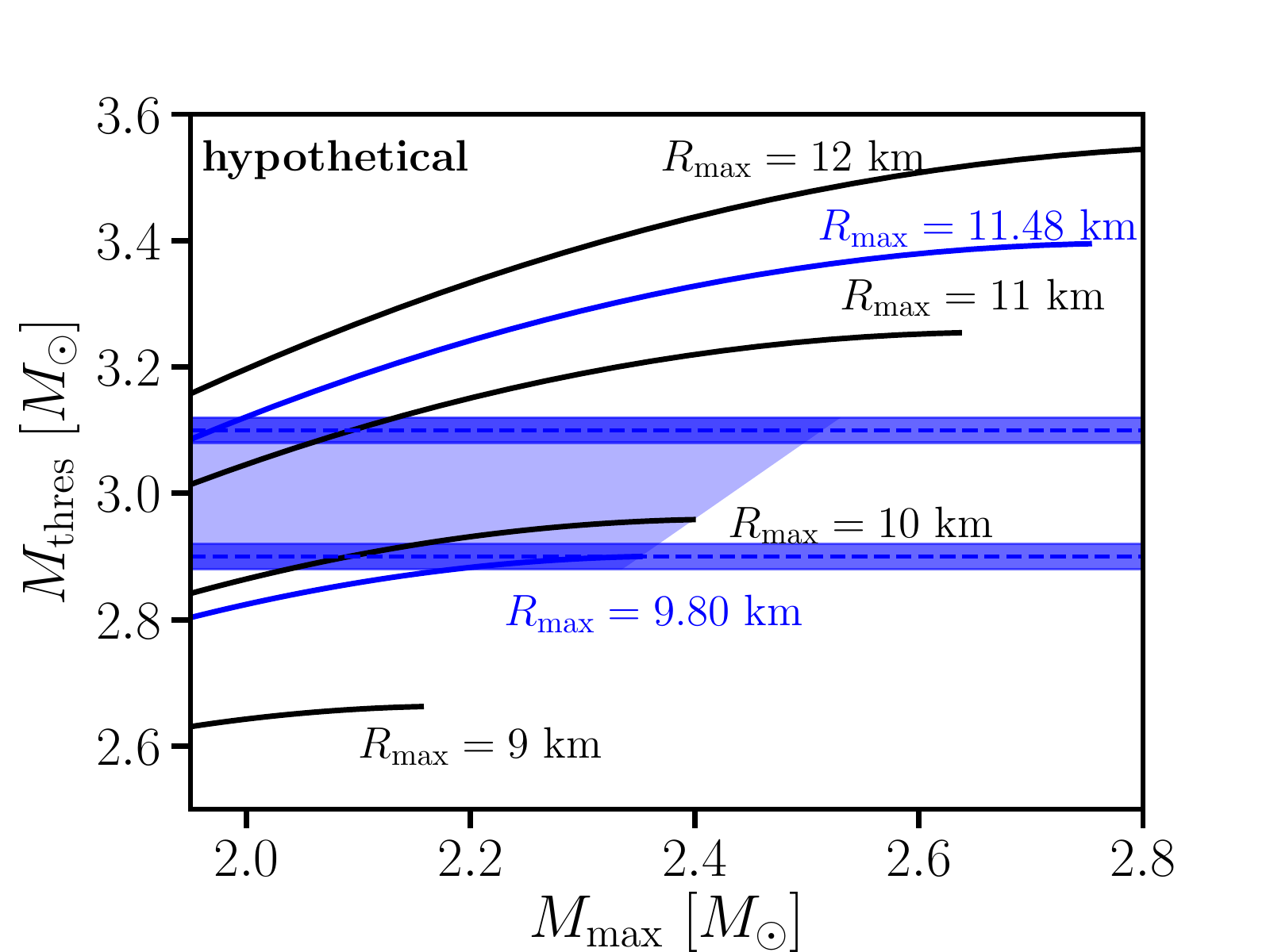}
\caption{Same as Fig.~\ref{fig:mthresmmax} (right panel). Dark blue bands display binary masses of hypothetical events with 2.9~$M_\odot$ resulting in a delayed collapse and 3.1~$M_\odot$ resulting in a prompt collapse. Viable NS properties are constrained to the light blue area.\label{fig:pcollapse}}
\end{figure}

(1) If an event with higher chirp mass than in GW170818 is detected and evidence for a delayed/no collapse is found, the lower bound on $M_\mathrm{thres}$ increases. The dark blue band in Fig.~\ref{fig:mthresmmax} shifts to higher $M_\mathrm{thres}$ and NS radii must be larger than implied by GW170817. This is sketched in Fig.~\ref{fig:pcollapse} for a hypothetical event with $M_\mathrm{tot}=(2.9\pm 0.02)~M_\odot$.

\begin{figure}[ht!]
\includegraphics[width=9cm]{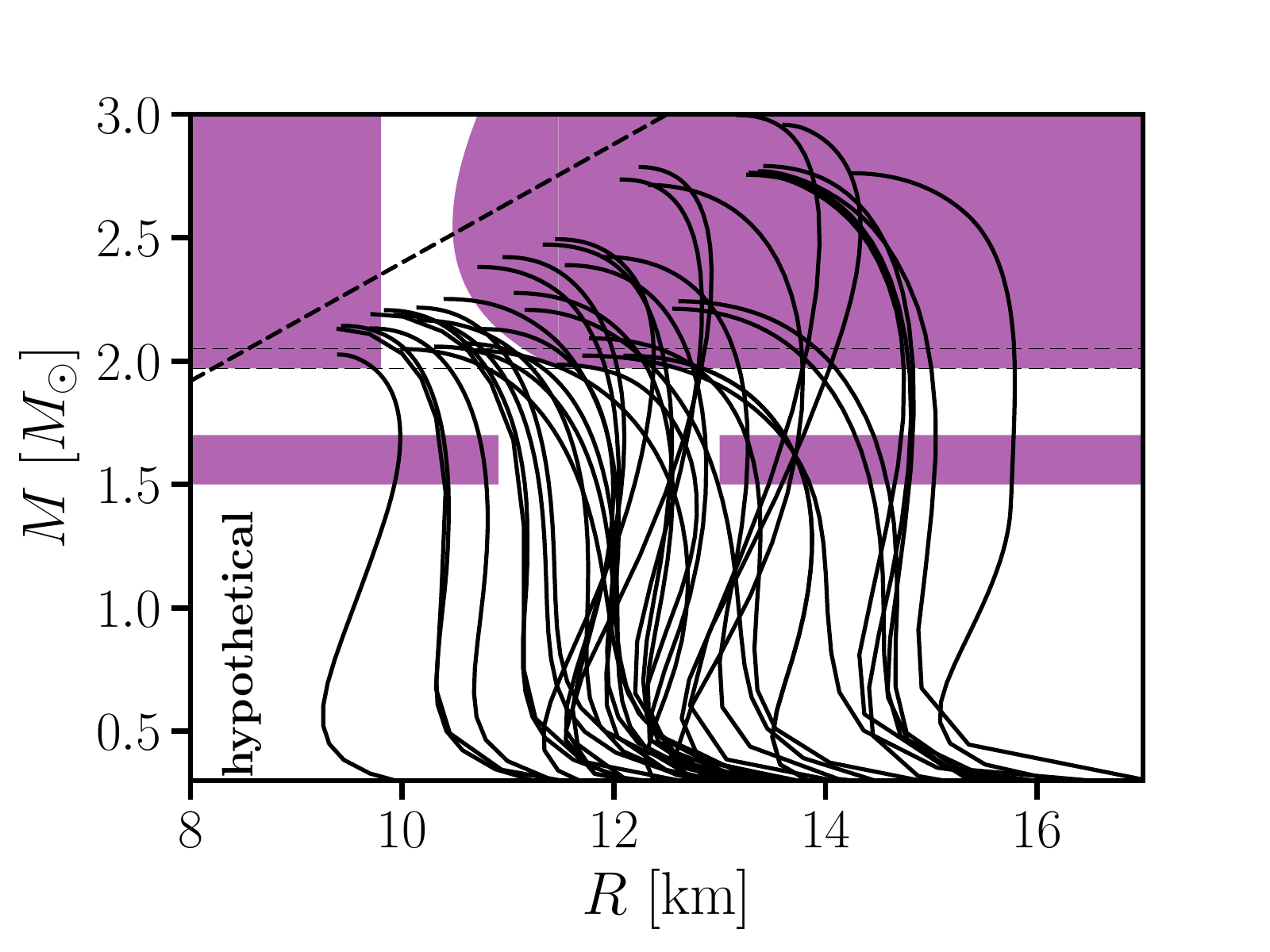}
\caption{Mass-radius relations of different EoSs with hypothetical exclusion regions (purple areas) from a delayed-collapse event with $M_\mathrm{tot}=2.9~M_\odot$ and a prompt-collapse event with $M_\mathrm{tot}=3.1~M_\odot$ employing the methods of this work (cf. Fig.~\ref{fig:pcollapse}).\label{fig:hypotov}}
\end{figure}

(2) If an event with a higher chirp mass than in GW170817 and a signature of a prompt collapse is observed, this will establish an upper bound on $M_\mathrm{thres}$. Figure~\ref{fig:pcollapse} shows this case for a hypothetical binary mass of 3.1~$M_\odot$. This measurement would imply an upper bound on NS radii, here $R_{16}\leq 13$~km and $R_\mathrm{max}\leq 11.48$~km, and an upper bound on $M_\mathrm{max}$ ($\sim2.5~M_\odot$ for this hypothetical case). These limits are visualized in Fig.~\ref{fig:hypotov}. The upper right exclusion region is given by the solution to $M_\mathrm{tot}=3.1~M_\odot=(-3.38\frac{G M_\mathrm{max}}{c^2 R_\mathrm{max}}+2.43)M_\mathrm{max}$ (Eq.~\ref{eq:mthrrmax}). %The constraints can become tighter if additional information from other measurements becomes available. For instance, if a new pulsar mass measurement increases the current lower limit on the maximum NS mass, the upper limit on NS radii will become even smaller. If another method constrains NSs to be smaller, the upper bound on $M_\mathrm{max}$ will become smaller. 
As more detections with different binary masses are made, $M_\mathrm{thres}$ will be constrained increasingly tighter from above and below. This will limit NS radii, i.e. $R_\mathrm{max}$ and $R_{1.6}$, and $M_\mathrm{max}$ to a relatively narrow range. $M_\mathrm{max}$ will be constrained from above and possibly determined with good accuracy if NS radii can be narrowed down by other even more accurate methods.

\begin{figure}[ht!]
\includegraphics[width=9cm]{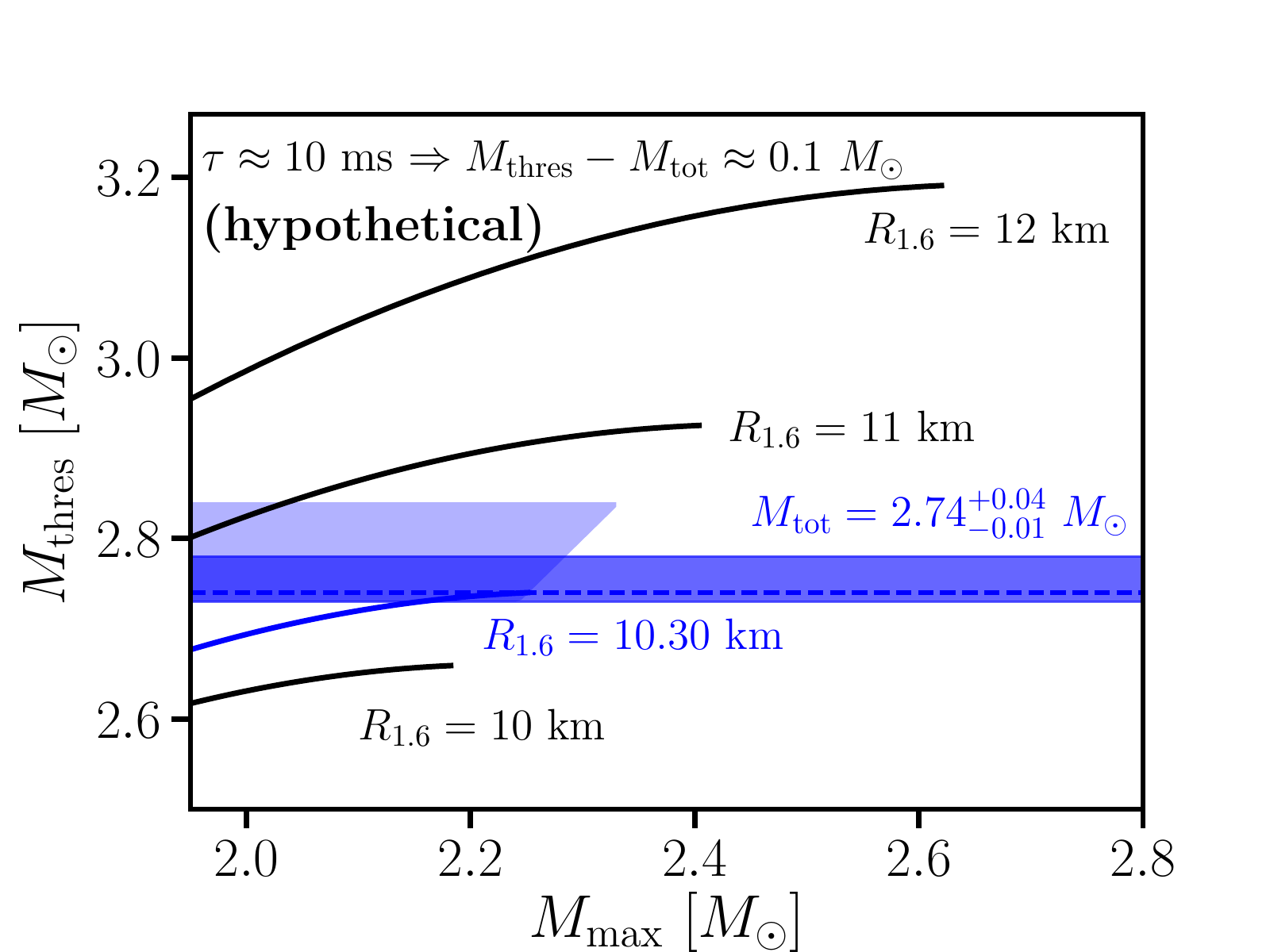}
\caption{Same as Fig.~\ref{fig:mthresmmax} (left panel) hypothetically assuming evidence for a remnant lifetime of $\tau \leq 10$~ms in an event like GW170817. NS properties $R_{1.6}$ and $M_\mathrm{max}$ would be constrained to the light blue area implying tight bounds on $R_{1.6}$.\label{fig:tauconstrained}}
\end{figure}

(3) Events with an upper bound on the remnant lifetime establish effectively an upper bound on $M_\mathrm{thres}$ with similar implications as in the previous scenario. This requires a better understanding of the exact dependence of the lifetime on binary masses and a reliable way to constrain the lifetime from observations, both of which can be achieved through improved numerical or analytic models. We sketch a hypothetical case in Fig.~\ref{fig:tauconstrained}.

%(3) If for a GW event the remnant life time can be constrained from above, it will approximately determine the proximity from the prompt collapse. This effectively establishes an upper bound on $M_\mathrm{thres}$ as in the previous scenario with the corresponding implications. This procedure requires more detailed numerical simulations to understand the exact dependence of the life time on binary masses and a solid  way to constrain the life time from observations. We sketch a hypothetical case in Fig.~\ref{fig:tauconstrained} assuming that in an event like GW170817 the life time was less than 10~ms. As discussed in Sect.~\ref{sec:discussion} one may then assume that a life time of $\tau \approx 10$~ms roughly implies $M_\mathrm{thres}-M_\mathrm{tot}\approx 0.15~M_\odot$. This hypothetical case would thus constrain NS radii from above and below, while $M_\mathrm{max}$ is bound from above (see light blue area in Fig.~\ref{fig:tauconstrained}).


\section{Conclusions}\label{sec:summary}
We introduce a new method to constrain NS radii and the maximum mass from GW observations of NS mergers and the observational distinction between a delayed and prompt collapse of the merger remnant. Based on the binary mass measurement of GW170817 and the well justified hypothesis of a delayed/no collapse in this event \citep[e.g.][]{Margalit2017,Metzger2017,Nicholl2017}, we show that the radius of a 1.6~$M_\odot$ NS must be larger than 
$10.68_{-0.04}^{+0.15}$~km and the radius of the maximum-mass configuration, $R_\mathrm{max}$, is larger than 
$9.60_{-0.03}^{+0.14}$~km. We stress the potential of future GW events. In particular, an event associated with a prompt collapse will constrain NS radii from above as well as the maximum mass $M_\mathrm{max}$ of nonrotating NSs. As the sensitivity of GW detectors increases, more events with more accurate mass measurements can be expected. Similarly, we anticipate a more robust identification of the collapse behavior as more electromagnetic counterparts are observed and increasingly better understood theoretically.

Our new method is particularly promising because it does not require higher SNRs of future GW events and is thus directly applicable to any new event within the era of current detectors for which the collapse behavior can be classified. It provides a robust, complimentary way of constraining the high-density EoS independent of efforts to measure finite-size effects during the late inspiral phase \citep{Faber2002,Flanagan2008,Read2013,DelPozzo2013,Abbott2017} or prospects to detect oscillations from the postmerger phase \citep{Bauswein2012,Bauswein2012a,Bauswein2014,Clark2014,Chatziioannou2017}. See e.g. \cite{Lawrence2015,Fryer2015,Margalit2017} for alternative methods to constrain $M_\mathrm{max}$.

Apart from the model-dependent interpretation of the electromagnetic emission our method only relies on binary mass measurements and empirical relations describing $M_\mathrm{thres}(M_\mathrm{max},R)$. Future calculations can further corroborate these relations for a larger sample of candidate EoSs and with more sophisticated models, although it seems unlikely that for instance a detailed incorporation of neutrinos or magnetic fields can have a significant influence on the relations for the threshold mass. We emphasize the simplicity and robustness of our constraints as a major advantage.

We demonstrated this robustness with the observation of GW170817 and its electromagnetic counterpart making conservative assumptions throughout, for instance by assuming an equal-mass merger. Future work should refine this first study and will yield stronger radius constraints. Specifically we refer to the inclusion of mass-ratio effects and additional information from limits on the remnant lifetime. %We will also extend this work to NS radii of other masses and will explore in detail other analytic expressions describing $M_\mathrm{thres}(M_\mathrm{max},R)$. 
As follow-up to this letter we will update our radius constraints following the methods described here as new measurements become available\footnote{Updated constraints will be published on http://wwwmpa.mpa-garching.mpg.de/$\sim$bauswein/radiusconstraint/}.

\begin{acknowledgements}
We thank K. Hebeler for EoS tables. A.B. acknowlegdes support by the Klaus Tschira Foundation. Partial support comes from COST actions GWverse CA16104 and PHAROS CA16214 and from the DAAD Germany-Greece grant ID 57340132. H.-T. J. acknowledges support by the European Research Council through grant ERC  AdG  341157-COCO2CASA  and  by  the  Deutsche Forschungsgemeinschaft through grants SFB 1258 and EXC 153 and thanks the Mainz Institute for Theoretical Physics.
\end{acknowledgements}

%\bibliography{refs}

\begin{thebibliography}{}
\expandafter\ifx\csname natexlab\endcsname\relax\def\natexlab#1{#1}\fi
\providecommand{\url}[1]{\href{#1}{#1}}

\bibitem[{Abbott {et~al.}(2017)Abbott, Abbott, Abbott, Acernese, Ackley, Adams,
  Adams, Addesso, Adhikari, Adya, Affeldt, Afrough, Agarwal, Agathos, Agatsuma,
  Aggarwal, Aguiar, Aiello, Ain, Ajith, Allen, Allen, Allocca, Altin, Amato,
  Ananyeva, Anderson, Anderson, Angelova, Antier, Appert, Arai, Araya, Areeda,
  Arnaud, Arun, Ascenzi, Ashton, Ast, Aston, Astone, Atallah, Aufmuth, Aulbert,
  AultONeal, Austin, Avila-Alvarez, Babak, Bacon, Bader, Bae, Bailes, Baker,
  Baldaccini, Ballardin, Ballmer, Banagiri, Barayoga, Barclay, Barish, Barker,
  Barkett, Barone, Barr, Barsotti, Barsuglia, Barta, Barthelmy, Bartlett,
  Bartos, Bassiri, Basti, Batch, Bawaj, Bayley, Bazzan, B\'ecsy, Beer, Bejger,
  Belahcene, Bell, Berger, Bergmann, Bernuzzi, Bero, Berry, Bersanetti,
  Bertolini, Betzwieser, Bhagwat, Bhandare, Bilenko, Billingsley, Billman,
  Birch, Birney, Birnholtz, Biscans, Biscoveanu, Bisht, Bitossi, Biwer,
  Bizouard, Blackburn, Blackman, Blair, Blair, Blair, Bloemen, Bock, Bode,
  Boer, Bogaert, Bohe, Bondu, Bonilla, Bonnand, Boom, Bork, Boschi, Bose,
  Bossie, Bouffanais, Bozzi, Bradaschia, Brady, Branchesi, Brau, Briant,
  Brillet, Brinkmann, Brisson, Brockill, Broida, Brooks, Brown, Brown, Brunett,
  Buchanan, Buikema, Bulik, Bulten, Buonanno, Buskulic, Buy, Byer, Cabero,
  Cadonati, Cagnoli, Cahillane, Calder\'on~Bustillo, Callister, Calloni, Camp,
  Canepa, Canizares, Cannon, Cao, Cao, Capano, Capocasa, Carbognani, Caride,
  Carney, Carullo, Casanueva~Diaz, Casentini, Caudill, Cavagli\`a, Cavalier,
  Cavalieri, Cella, Cepeda, Cerd\'a-Dur\'an, Cerretani, Cesarini, Chamberlin,
  Chan, Chao, Charlton, Chase, Chassande-Mottin, Chatterjee, Chatziioannou,
  Cheeseboro, Chen, Chen, Chen, Cheng, Chia, Chincarini, Chiummo, Chmiel, Cho,
  Cho, Chow, Christensen, Chu, Chua, Chua, Chung, Chung, Ciani, Ciolfi,
  Cirelli, Cirone, Clara, Clark, Clearwater, Cleva, Cocchieri, Coccia, Cohadon,
  Cohen, Colla, Collette, Cominsky, Constancio, Conti, Cooper, Corban, Corbitt,
  Cordero-Carri\'on, Corley, Cornish, Corsi, Cortese, Costa, Coughlin,
  Coughlin, Coulon, Countryman, Couvares, Covas, Cowan, Coward, Cowart, Coyne,
  Coyne, Creighton, Creighton, Cripe, Crowder, Cullen, Cumming, Cunningham,
  Cuoco, Dal~Canton, D\'alya, Danilishin, D'Antonio, Danzmann, Dasgupta,
  Da~Silva~Costa, Dattilo, Dave, Davier, Davis, Daw, Day, De, DeBra, Degallaix,
  De~Laurentis, Del\'eglise, Del~Pozzo, Demos, Denker, Dent, De~Pietri,
  Dergachev, De~Rosa, DeRosa, De~Rossi, DeSalvo, de~Varona, Devenson,
  Dhurandhar, D\'{\i}az, Dietrich, Di~Fiore, Di~Giovanni, Di~Girolamo,
  Di~Lieto, Di~Pace, Di~Palma, Di~Renzo, Doctor, Dolique, Donovan, Dooley,
  Doravari, Dorrington, Douglas, Dovale~\'Alvarez, Downes, Drago,
  Dreissigacker, Driggers, Du, Ducrot, Dudi, Dupej, Dwyer, Edo, Edwards,
  Effler, Eggenstein, Ehrens, Eichholz, Eikenberry, Eisenstein, Essick,
  Estevez, Etienne, Etzel, Evans, Evans, Factourovich, Fafone, Fair, Fairhurst,
  Fan, Farinon, Farr, Farr, Fauchon-Jones, Favata, Fays, Fee, Fehrmann, Feicht,
  Fejer, Fernandez-Galiana, Ferrante, Ferreira, Ferrini, Fidecaro, Finstad,
  Fiori, Fiorucci, Fishbach, Fisher, Fitz-Axen, Flaminio, Fletcher, Fong, Font,
  Forsyth, Forsyth, Fournier, Frasca, Frasconi, Frei, Freise, Frey, Frey,
  Fries, Fritschel, Frolov, Fulda, Fyffe, Gabbard, Gadre, Gaebel, Gair,
  Gammaitoni, Ganija, Gaonkar, Garcia-Quiros, Garufi, Gateley, Gaudio, Gaur,
  Gayathri, Gehrels, Gemme, Genin, Gennai, George, George, Gergely, Germain,
  Ghonge, Ghosh, Ghosh, Ghosh, Giaime, Giardina, Giazotto, Gill, Glover, Goetz,
  Goetz, Gomes, Goncharov, Gonz\'alez, Gonzalez~Castro, Gopakumar, Gorodetsky,
  Gossan, Gosselin, Gouaty, Grado, Graef, Granata, Grant, Gras, Gray, Greco,
  Green, Gretarsson, Groot, Grote, Grunewald, Gruning, Guidi, Guo, Gupta,
  Gupta, Gushwa, Gustafson, Gustafson, Halim, Hall, Hall, Hamilton, Hammond,
  Haney, Hanke, Hanks, Hanna, Hannam, Hannuksela, Hanson, Hardwick, Harms,
  Harry, Harry, Hart, Haster, Haughian, Healy, Heidmann, Heintze, Heitmann,
  Hello, Hemming, Hendry, Heng, Hennig, Heptonstall, Heurs, Hild, Hinderer, Ho,
  Hoak, Hofman, Holt, Holz, Hopkins, Horst, Hough, Houston, Howell, Hreibi, Hu,
  Huerta, Huet, Hughey, Husa, Huttner, Huynh-Dinh, Indik, Inta, Intini, Isa,
  Isac, Isi, Iyer, Izumi, Jacqmin, Jani, Jaranowski, Jawahar,
  Jim\'enez-Forteza, Johnson, Johnson-McDaniel, Jones, Jones, Jonker, Ju,
  Junker, Kalaghatgi, Kalogera, Kamai, Kandhasamy, Kang, Kanner, Kapadia,
  Karki, Karvinen, Kasprzack, Kastaun, Katolik, Katsavounidis, Katzman, Kaufer,
  Kawabe, K\'ef\'elian, Keitel, Kemball, Kennedy, Kent, Key, Khalili, Khan,
  Khan, Khan, Khazanov, Kijbunchoo, Kim, Kim, Kim, Kim, Kim, Kim, Kimbrell,
  King, King, Kinley-Hanlon, Kirchhoff, Kissel, Kleybolte, Klimenko, Knowles,
  Koch, Koehlenbeck, Koley, Kondrashov, Kontos, Korobko, Korth, Kowalska,
  Kozak, Kr\"amer, Kringel, Krishnan, Kr\'olak, Kuehn, Kumar, Kumar, Kumar,
  Kuo, Kutynia, Kwang, Lackey, Lai, Landry, Lang, Lange, Lantz, Lanza, Larson,
  Lartaux-Vollard, Lasky, Laxen, Lazzarini, Lazzaro, Leaci, Leavey, Lee, Lee,
  Lee, Lee, Lee, Lehmann, Lenon, Leon, Leonardi, Leroy, Letendre, Levin, Li,
  Linker, Littenberg, Liu, Liu, Lo, Lockerbie, London, Lord, Lorenzini,
  Loriette, Lormand, Losurdo, Lough, Lousto, Lovelace, L\"uck, Lumaca,
  Lundgren, Lynch, Ma, Macas, Macfoy, Machenschalk, MacInnis, Macleod, Maga\~na
  Hernandez, Maga\~na Sandoval, Maga\~na Zertuche, Magee, Majorana, Maksimovic,
  Man, Mandic, Mangano, Mansell, Manske, Mantovani, Marchesoni, Marion,
  M\'arka, M\'arka, Markakis, Markosyan, Markowitz, Maros, Marquina, Marsh,
  Martelli, Martellini, Martin, Martin, Martynov, Marx, Mason, Massera,
  Masserot, Massinger, Masso-Reid, Mastrogiovanni, Matas, Matichard, Matone,
  Mavalvala, Mazumder, McCarthy, McClelland, McCormick, McCuller, McGuire,
  McIntyre, McIver, McManus, McNeill, McRae, McWilliams, Meacher, Meadors,
  Mehmet, Meidam, Mejuto-Villa, Melatos, Mendell, Mercer, Merilh, Merzougui,
  Meshkov, Messenger, Messick, Metzdorff, Meyers, Miao, Michel, Middleton,
  Mikhailov, Milano, Miller, Miller, Miller, Millhouse, Milovich-Goff,
  Minazzoli, Minenkov, Ming, Mishra, Mitra, Mitrofanov, Mitselmakher,
  Mittleman, Moffa, Moggi, Mogushi, Mohan, Mohapatra, Molina, Montani, Moore,
  Moraru, Moreno, Morisaki, Morriss, Mours, Mow-Lowry, Mueller, Muir,
  Mukherjee, Mukherjee, Mukherjee, Mukund, Mullavey, Munch, Mu\~niz, Muratore,
  Murray, Nagar, Napier, Nardecchia, Naticchioni, Nayak, Neilson, Nelemans,
  Nelson, Nery, Neunzert, Nevin, Newport, Newton, Ng, Nguyen, Nguyen, Nichols,
  Nielsen, Nissanke, Nitz, Noack, Nocera, Nolting, North, Nuttall, Oberling,
  O'Dea, Ogin, Oh, Oh, Ohme, Okada, Oliver, Oppermann, Oram, O'Reilly,
  Ormiston, Ortega, O'Shaughnessy, Ossokine, Ottaway, Overmier, Owen, Pace,
  Page, Page, Pai, Pai, Palamos, Palashov, Palomba, Pal-Singh, Pan, Pan, Pang,
  Pang, Pankow, Pannarale, Pant, Paoletti, Paoli, Papa, Parida, Parker,
  Pascucci, Pasqualetti, Passaquieti, Passuello, Patil, Patricelli, Pearlstone,
  Pedraza, Pedurand, Pekowsky, Pele, Penn, Perez, Perreca, Perri, Pfeiffer,
  Phelps, Piccinni, Pichot, Piergiovanni, Pierro, Pillant, Pinard, Pinto,
  Pirello, Pitkin, Poe, Poggiani, Popolizio, Porter, Post, Powell, Prasad,
  Pratt, Pratten, Predoi, Prestegard, Prijatelj, Principe, Privitera, Prix,
  Prodi, Prokhorov, Puncken, Punturo, Puppo, P\"urrer, Qi, Quetschke, Quintero,
  Quitzow-James, Raab, Rabeling, Radkins, Raffai, Raja, Rajan, Rajbhandari,
  Rakhmanov, Ramirez, Ramos-Buades, Rapagnani, Raymond, Razzano, Read,
  Regimbau, Rei, Reid, Reitze, Ren, Reyes, Ricci, Ricker, Rieger, Riles, Rizzo,
  Robertson, Robie, Robinet, Rocchi, Rolland, Rollins, Roma, Romano, Romano,
  Romel, Romie, Rosi\ifmmode~\acute{n}\else \'{n}\fi{}ska, Ross, Rowan,
  R\"udiger, Ruggi, Rutins, Ryan, Sachdev, Sadecki, Sadeghian, Sakellariadou,
  Salconi, Saleem, Salemi, Samajdar, Sammut, Sampson, Sanchez, Sanchez,
  Sanchis-Gual, Sandberg, Sanders, Sassolas, Sathyaprakash, Saulson, Sauter,
  Savage, Sawadsky, Schale, Scheel, Scheuer, Schmidt, Schmidt, Schnabel,
  Schofield, Sch\"onbeck, Schreiber, Schuette, Schulte, Schutz, Schwalbe,
  Scott, Scott, Seidel, Sellers, Sengupta, Sentenac, Sequino, Sergeev,
  Shaddock, Shaffer, Shah, Shahriar, Shaner, Shao, Shapiro, Shawhan, Sheperd,
  Shoemaker, Shoemaker, Siellez, Siemens, Sieniawska, Sigg, Silva, Singer,
  Singh, Singhal, Sintes, Slagmolen, Smith, Smith, Smith, Somala, Son,
  Sonnenberg, Sorazu, Sorrentino, Souradeep, Spencer, Srivastava, Staats,
  Staley, Steinke, Steinlechner, Steinlechner, Steinmeyer, Stevenson, Stone,
  Stops, Strain, Stratta, Strigin, Strunk, Sturani, Stuver, Summerscales, Sun,
  Sunil, Suresh, Sutton, Swinkels, Szczepa\ifmmode~\acute{n}\else
  \'{n}\fi{}czyk, Tacca, Tait, Talbot, Talukder, Tanner, T\'apai, Taracchini,
  Tasson, Taylor, Taylor, Tewari, Theeg, Thies, Thomas, Thomas, Thomas, Thorne,
  Thorne, Thrane, Tiwari, Tiwari, Tokmakov, Toland, Tonelli, Tornasi,
  Torres-Forn\'e, Torrie, T\"oyr\"a, Travasso, Traylor, Trinastic, Tringali,
  Trozzo, Tsang, Tse, Tso, Tsukada, Tsuna, Tuyenbayev, Ueno, Ugolini,
  Unnikrishnan, Urban, Usman, Vahlbruch, Vajente, Valdes, Vallisneri, van
  Bakel, van Beuzekom, van~den Brand, Van Den~Broeck, Vander-Hyde, van~der
  Schaaf, van Heijningen, van Veggel, Vardaro, Varma, Vass, Vas\'uth, Vecchio,
  Vedovato, Veitch, Veitch, Venkateswara, Venugopalan, Verkindt, Vetrano,
  Vicer\'e, Viets, Vinciguerra, Vine, Vinet, Vitale, Vo, Vocca, Vorvick,
  Vyatchanin, Wade, Wade, Wade, Walet, Walker, Wallace, Walsh, Wang, Wang,
  Wang, Wang, Wang, Ward, Warner, Was, Watchi, Weaver, Wei, Weinert, Weinstein,
  Weiss, Wen, Wessel, We\ss{}els, Westerweck, Westphal, Wette, Whelan,
  Whitcomb, Whiting, Whittle, Wilken, Williams, Williams, Williamson, Willis,
  Willke, Wimmer, Winkler, Wipf, Wittel, Woan, Woehler, Wofford, Wong, Worden,
  Wright, Wu, Wysocki, Xiao, Yamamoto, Yancey, Yang, Yap, Yazback, Yu, Yu,
  Yvert, Zadro\ifmmode~\dot{z}\else \.{z}\fi{}ny, Zanolin, Zelenova, Zendri,
  Zevin, Zhang, Zhang, Zhang, Zhang, Zhao, Zhou, Zhou, Zhu, Zhu, Zimmerman,
  Zucker, \& Zweizig}]{Abbott2017}
Abbott, B.~P., Abbott, R., Abbott, T.~D., {et~al.} 2017, Phys. Rev. Lett., 119,
  161101

\bibitem[{{Antoniadis} \& {et al.}(2013)}]{Antoniadis2013}
{Antoniadis}, J., \& {et al.} 2013, Science, 340, 448

\bibitem[{Baiotti {et~al.}(2008)Baiotti, Giacomazzo, \& Rezzolla}]{Baiotti2008}
Baiotti, L., Giacomazzo, B., \& Rezzolla, L. 2008, \prd, 78, 084033

\bibitem[{{Bauswein} {et~al.}(2013{\natexlab{a}}){Bauswein}, {Baumgarte}, \&
  {Janka}}]{Bauswein2013}
{Bauswein}, A., {Baumgarte}, T.~W., \& {Janka}, H.-T. 2013{\natexlab{a}}, \prl,
  111, 131101

\bibitem[{{Bauswein} {et~al.}(2013{\natexlab{b}}){Bauswein}, {Goriely}, \&
  {Janka}}]{Bauswein2013a}
{Bauswein}, A., {Goriely}, S., \& {Janka}, H.-T. 2013{\natexlab{b}}, \apj, 773,
  78

\bibitem[{{Bauswein} \& {Janka}(2012)}]{Bauswein2012}
{Bauswein}, A., \& {Janka}, H.-T. 2012, \prl, 108, 011101

\bibitem[{{Bauswein} {et~al.}(2012){Bauswein}, {Janka}, {Hebeler}, \&
  {Schwenk}}]{Bauswein2012a}
{Bauswein}, A., {Janka}, H.-T., {Hebeler}, K., \& {Schwenk}, A. 2012, \prd, 86,
  063001

\bibitem[{{Bauswein} \& {Stergioulas}(2017)}]{Bauswein2017}
{Bauswein}, A., \& {Stergioulas}, N. 2017, \mnras, 471, 4956

\bibitem[{{Bauswein} {et~al.}(2014){Bauswein}, {Stergioulas}, \&
  {Janka}}]{Bauswein2014}
{Bauswein}, A., {Stergioulas}, N., \& {Janka}, H.-T. 2014, \prd, 90, 023002

\bibitem[{{Bauswein} {et~al.}(2016){Bauswein}, {Stergioulas}, \&
  {Janka}}]{Bauswein2016}
---. 2016, European Physical Journal A, 52, 56

\bibitem[{{Chatziioannou} {et~al.}(2017){Chatziioannou}, {Clark}, {Bauswein},
  {Littenberg}, \& {Cornish}}]{Chatziioannou2017}
{Chatziioannou}, K., {Clark}, J.~A., {Bauswein}, A.~{Millhouse}, M.,
  {Littenberg}, T., \& {Cornish}, N. 2017, ArXiv e-prints, arXiv:1711.00040

\bibitem[{Chornock {et~al.}(2017)Chornock, Berger, Kasen, Cowperthwaite,
  Nicholl, Villar, Alexander, Blanchard, Eftekhari, Fong, Margutti, Williams,
  Annis, Brout, Brown, Chen, Drout, Foley, Frieman, Fryer, Holz, Matheson,
  Metzger, Quataert, Rest, Sako, Scolnic, Smith, \&
  Soares-Santos}]{Chornock2017}
Chornock, R., Berger, E., Kasen, D., {et~al.} 2017, \apjl, 848, L18

\bibitem[{{Clark} {et~al.}(2014){Clark}, {Bauswein}, {Cadonati}, {Janka},
  {Pankow}, \& {Stergioulas}}]{Clark2014}
{Clark}, J., {Bauswein}, A., {Cadonati}, L., {et~al.} 2014, \prd, 90, 062004

\bibitem[{Cowperthwaite {et~al.}(2017)Cowperthwaite, Berger, Villar, Metzger,
  Nicholl, Chornock, Blanchard, Fong, Margutti, Soares-Santos, Alexander,
  Allam, Annis, Brout, Brown, Butler, Chen, Diehl, Doctor, Drout, Eftekhari,
  Farr, Finley, Foley, Frieman, Fryer, Garc{\'\i}a-Bellido, Gill, Guillochon,
  Herner, Holz, Kasen, Kessler, Marriner, Matheson, E.~H.~Neilsen, Quataert,
  Palmese, Rest, Sako, Scolnic, Smith, Tucker, Williams, Balbinot, Carlin,
  Cook, Durret, Li, Lopes, Louren{\c c}o, Marshall, Medina, Muir, Mu{\~n}oz,
  Sauseda, Schlegel, Secco, Vivas, \& et~al. (85 additional authors~not
  shown)}]{Cowperthwaite2017}
Cowperthwaite, P.~S., Berger, E., Villar, V.~A., {et~al.} 2017, \apjl, 848, L17

\bibitem[{{Del Pozzo} {et~al.}(2013){Del Pozzo}, {Li}, {Agathos}, {Van Den
  Broeck}, \& {Vitale}}]{DelPozzo2013}
{Del Pozzo}, W., {Li}, T.~G.~F., {Agathos}, M., {Van Den Broeck}, C., \&
  {Vitale}, S. 2013, Physical Review Letters, 111, 071101

\bibitem[{Drout {et~al.}(2017)Drout, Piro, Shappee, Kilpatrick, Simon,
  Contreras, Coulter, Foley, Siebert, Morrell, Boutsia, Di~Mille, Holoien,
  Kasen, Kollmeier, Madore, Monson, Murguia-Berthier, Pan, Prochaska,
  Ramirez-Ruiz, Rest, Adams, Alatalo, Ba{\~n}ados, Baughman, Beers, Bernstein,
  Bitsakis, Campillay, Hansen, Higgs, Ji, Maravelias, Marshall, Moni~Bidin,
  Prieto, Rasmussen, Rojas-Bravo, Strom, Ulloa, Vargas-Gonz{\'a}lez, Wan, \&
  Whitten}]{Drout2017}
Drout, M.~R., Piro, A.~L., Shappee, B.~J., {et~al.} 2017, ArXiv e-prints,
  arXiv:1710.05443

\bibitem[{{Faber} {et~al.}(2002){Faber}, {Grandcl{\'e}ment}, {Rasio}, \&
  {Taniguchi}}]{Faber2002}
{Faber}, J.~A., {Grandcl{\'e}ment}, P., {Rasio}, F.~A., \& {Taniguchi}, K.
  2002, Physical Review Letters, 89, 231102

\bibitem[{{Fern{\'a}ndez} \& {Metzger}(2013)}]{Fernandez2013}
{Fern{\'a}ndez}, R., \& {Metzger}, B.~D. 2013, \mnras, 435, 502

\bibitem[{{Flanagan} \& {Hinderer}(2008)}]{Flanagan2008}
{Flanagan}, {\'E}.~{\'E}., \& {Hinderer}, T. 2008, \prd, 77, 021502

\bibitem[{{Fryer} {et~al.}(2015){Fryer}, {Belczynski}, {Ramirez-Ruiz},
  {Rosswog}, {Shen}, \& {Steiner}}]{Fryer2015}
{Fryer}, C.~L., {Belczynski}, K., {Ramirez-Ruiz}, E., {et~al.} 2015, \apj, 812,
  24

\bibitem[{Hebeler {et~al.}(2013)Hebeler, Lattimer, Pethick, \&
  Schwenk}]{Hebeler2013}
Hebeler, K., Lattimer, J.~M., Pethick, C.~J., \& Schwenk, A. 2013, \apj, 773,
  11

\bibitem[{{Hotokezaka} {et~al.}(2013){Hotokezaka}, {Kiuchi}, {Kyutoku},
  {Okawa}, {Sekiguchi}, {Shibata}, \& {Taniguchi}}]{Hotokezaka2013}
{Hotokezaka}, K., {Kiuchi}, K., {Kyutoku}, K., {et~al.} 2013, \prd, 87, 024001

\bibitem[{{Hotokezaka} {et~al.}(2011){Hotokezaka}, {Kyutoku}, {Okawa},
  {Shibata}, \& {Kiuchi}}]{Hotokezaka2011}
{Hotokezaka}, K., {Kyutoku}, K., {Okawa}, H., {Shibata}, M., \& {Kiuchi}, K.
  2011, \prd, 83, 124008

\bibitem[{{Just} {et~al.}(2015){Just}, {Bauswein}, {Pulpillo}, {Goriely}, \&
  {Janka}}]{Just2015}
{Just}, O., {Bauswein}, A., {Pulpillo}, R.~A., {Goriely}, S., \& {Janka}, H.-T.
  2015, \mnras, 448, 541

\bibitem[{Kasen {et~al.}(2017)Kasen, Metzger, Barnes, Quataert, \&
  Ramirez-Ruiz}]{Kasen2017}
Kasen, D., Metzger, B., Barnes, J., Quataert, E., \& Ramirez-Ruiz, E. 2017,
  \nat, 1710.05463

\bibitem[{Kasliwal {et~al.}(2017)Kasliwal, Nakar, Singer, Kaplan, Cook,
  Van~Sistine, Lau, Fremling, Gottlieb, Jencson, Adams, Feindt, Hotokezaka,
  Ghosh, Perley, Yu, Piran, Allison, Anupama, Balasubramanian, Bannister,
  Bally, Barnes, Barway, Bellm, Bhalerao, Bhattacharya, Blagorodnova, Bloom,
  Brady, Cannella, Chatterjee, Cenko, Cobb, Copperwheat, Corsi, De, Dobie,
  Emery, Evans, Fox, Frail, Frohmaier, Goobar, Hallinan, Harrison, Helou,
  Hinderer, Ho, Horesh, Ip, Itoh, Kasen, Kim, Kuin, Kupfer, Lynch, Madsen,
  Mazzali, Miller, Mooley, Murphy, Ngeow, Nichols, Nissanke, Nugent, Ofek, Qi,
  Quimby, Rosswog, Rusu, Sadler, Schmidt, Sollerman, Steele, Williamson, Xu,
  Yan, Yatsu, Zhang, \& Zhao}]{Kasliwal2017}
Kasliwal, M.~M., Nakar, E., Singer, L.~P., {et~al.} 2017, ArXiv e-prints,
  arXiv:1710.05436

\bibitem[{Kilpatrick {et~al.}(2017)Kilpatrick, Foley, Kasen, Murguia-Berthier,
  Ramirez-Ruiz, Coulter, Drout, Piro, Shappee, Boutsia, Contreras, Di~Mille,
  Madore, Morrell, Pan, Prochaska, Rest, Rojas-Bravo, Siebert, Simon, \&
  Ulloa}]{Kilpatrick2017}
Kilpatrick, C.~D., Foley, R.~J., Kasen, D., {et~al.} 2017, ArXiv e-prints,
  arXiv:1710.05434

\bibitem[{{Koranda} {et~al.}(1997){Koranda}, {Stergioulas}, \&
  {Friedman}}]{Koranda1997}
{Koranda}, S., {Stergioulas}, N., \& {Friedman}, J.~L. 1997, \apj, 488, 799

\bibitem[{{Lattimer} \& {Prakash}(2016)}]{Lattimer2016}
{Lattimer}, J.~M., \& {Prakash}, M. 2016, \physrep, 621, 127

\bibitem[{{Lawrence} {et~al.}(2015){Lawrence}, {Tervala}, {Bedaque}, \&
  {Miller}}]{Lawrence2015}
{Lawrence}, S., {Tervala}, J.~G., {Bedaque}, P.~F., \& {Miller}, M.~C. 2015,
  \apj, 808, 186

\bibitem[{{LIGO Scientific Collaboration} {et~al.}(2017){LIGO Scientific
  Collaboration}, {Virgo Collaboration}, \& et~al.}]{Multi2017}
{LIGO Scientific Collaboration}, {Virgo Collaboration}, \& et~al. 2017, \apjl,
  848, L12

\bibitem[{Lippuner {et~al.}(2017)Lippuner, Fern{\'a}ndez, Roberts, Foucart,
  Kasen, Metzger, \& Ott}]{Lippuner2017}
Lippuner, J., Fern{\'a}ndez, R., Roberts, L.~F., {et~al.} 2017, \mnras, 472,
  904

\bibitem[{{Margalit} \& {Metzger}(2017)}]{Margalit2017}
{Margalit}, B., \& {Metzger}, B. 2017, ArXiv e-prints, arXiv:1710.05938v1

\bibitem[{{Metzger}(2017)}]{Metzger2017}
{Metzger}, B.~D. 2017, ArXiv e-prints, arXiv:1710.05931v1

\bibitem[{{Metzger} \& {Fern{\'a}ndez}(2014)}]{Metzger2014}
{Metzger}, B.~D., \& {Fern{\'a}ndez}, R. 2014, \mnras, 441, 3444

\bibitem[{{Metzger} {et~al.}(2010){Metzger}, {Mart{\'{\i}}nez-Pinedo},
  {Darbha}, {Quataert}, {Arcones}, {Kasen}, {Thomas}, {Nugent}, {Panov}, \&
  {Zinner}}]{Metzger2010}
{Metzger}, B.~D., {Mart{\'{\i}}nez-Pinedo}, G., {Darbha}, S., {et~al.} 2010,
  \mnras, 406, 2650

\bibitem[{Nicholl {et~al.}(2017)Nicholl, Berger, Kasen, Metzger, Elias,
  Briceno, Alexander, Blanchard, Chornock, Cowperthwaite, Eftekhari, Fong,
  Margutti, Villar, Williams, Brown, Annis, Bahramian, Brout, Brown, Chen,
  Clemens, Dennihy, Dunlap, Holz, Marchesini, Massaro, Moskowitz, Pelisoli,
  Rest, Ricci, Sako, Soares-Santos, \& Strader}]{Nicholl2017}
Nicholl, M., Berger, E., Kasen, D., {et~al.} 2017, \apjl, 848, L18

\bibitem[{Oechslin {et~al.}(2007)Oechslin, Janka, \& Marek}]{Oechslin2007}
Oechslin, R., Janka, H.-T., \& Marek, A. 2007, \aap, 467, 395

\bibitem[{{Oertel} {et~al.}(2017){Oertel}, {Hempel}, {Kl{\"a}hn}, \&
  {Typel}}]{Oertel2017}
{Oertel}, M., {Hempel}, M., {Kl{\"a}hn}, T., \& {Typel}, S. 2017, Reviews of
  Modern Physics, 89, 015007

\bibitem[{{{\"O}zel} \& {Freire}(2016)}]{Oezel2016}
{{\"O}zel}, F., \& {Freire}, P. 2016, \araa, 54, 401

\bibitem[{Perego {et~al.}(2014)Perego, Rosswog, Cabez{\'o}n, Korobkin,
  K{\"a}ppeli, Arcones, \& Liebend{\"o}rfer}]{Perego2014}
Perego, A., Rosswog, S., Cabez{\'o}n, R.~M., {et~al.} 2014, \mnras, 443, 3134

\bibitem[{{Read} {et~al.}(2013){Read}, {Baiotti}, {Creighton}, {Friedman},
  {Giacomazzo}, {Kyutoku}, {Markakis}, {Rezzolla}, {Shibata}, \&
  {Taniguchi}}]{Read2013}
{Read}, J.~S., {Baiotti}, L., {Creighton}, J.~D.~E., {et~al.} 2013, \prd, 88,
  044042

\bibitem[{Sekiguchi {et~al.}(2016)Sekiguchi, Kiuchi, Kyutoku, Shibata, \&
  Taniguchi}]{Sekiguchi2016}
Sekiguchi, Y., Kiuchi, K., Kyutoku, K., Shibata, M., \& Taniguchi, K. 2016,
  \prd, 93, 124046

\bibitem[{{Shibata}(2005)}]{Shibata2005}
{Shibata}, M. 2005, \prl, 94, 201101

\bibitem[{Siegel {et~al.}(2014)Siegel, Ciolfi, \& Rezzolla}]{Siegel2014}
Siegel, D.~M., Ciolfi, R., \& Rezzolla, L. 2014, \apjl, 785, L6

\bibitem[{Smartt {et~al.}(2017)Smartt, Chen, Jerkstrand, Coughlin, Kankare,
  Sim, Fraser, Inserra, Maguire, Chambers, Huber, Kruhler, Leloudas, Magee,
  Shingles, Smith, Young, Tonry, Kotak, Gal-Yam, Lyman, Homan, Agliozzo,
  Anderson, Ashall, Barbarino, Bauer, Berton, Botticella, Bulla, Bulger,
  Cannizzaro, Cano, Cartier, Cikota, Clark, De~Cia, Della~Valle, Denneau,
  Dennefeld, Dessart, Dimitriadis, Elias-Rosa, Firth, Flewelling, Flors,
  Franckowiak, Frohmaier, Galbany, Gonzalez-Gaitan, Greiner, Gromadzki,
  Nicuesa~Guelbenzu, Gutierrez, Hamanowicz, Hanlon, Harmanen, Heintz, Heinze,
  Hernandez, Hodgkin, Hook, Izzo, James, Jonker, Kerzendorf, Klose,
  Kostrzewa-Rutkowska, Kowalski, Kromer, Kuncarayakti, Lawrence, Lowe, Magnier,
  Manulis, Martin-Carrillo, Mattila, McBrien, Muller, Nordin, O'Neill, Onori,
  Palmerio, Pastorello, Patat, Pignata, Podsiadlowski, Pumo, Prentice, Rau,
  Razza, Rest, Reynolds, Roy, Ruiter, Rybicki, Salmon, Schady, Schultz,
  Schweyer, Seitenzahl, Smith, Sollerman, Stalder, Stubbs, Sullivan, Szegedi,
  Taddia, Taubenberger, Terreran, van Soelen, Vos, Wainscoat, Walton, Waters,
  Weiland, Willman, Wiseman, Wright, Wyrzykowski, \& Yaron}]{Smartt2017}
Smartt, S.~J., Chen, T.-W., Jerkstrand, A., {et~al.} 2017, ArXiv e-prints,
  arXiv:1710.05841

\bibitem[{Takami {et~al.}(2014)Takami, Rezzolla, \& Baiotti}]{Takami2014}
Takami, K., Rezzolla, L., \& Baiotti, L. 2014, Physical Review Letters, 113,
  091104

\bibitem[{Tanaka {et~al.}(2017)Tanaka, Utsumi, Mazzali, Tominaga, Yoshida,
  Sekiguchi, Morokuma, Motohara, Ohta, Kawabata, Abe, Aoki, Asakura, Baar,
  Barway, Bond, Doi, Fujiyoshi, Furusawa, Honda, Itoh, Kawabata, Kawai, Kim,
  Lee, Miyazaki, Morihana, Nagashima, Nagayama, Nakaoka, Nakata, Ohsawa,
  Ohshima, Okita, Saito, Sumi, Tajitsu, Takahashi, Takayama, Tamura, Tanaka,
  Terai, Tristram, Yasuda, \& Zenko}]{Tanaka2017}
Tanaka, M., Utsumi, Y., Mazzali, P.~A., {et~al.} 2017, published in PASJ,
  1710.05850

\bibitem[{Tanvir {et~al.}(2017)Tanvir, Levan, Gonzalez-Fernandez, Korobkin,
  Mandel, Rosswog, Hjorth, D'Avanzo, Fruchter, Fryer, Kangas, Milvang-Jensen,
  Rosetti, Steeghs, Wollaeger, Cano, Copperwheat, Covino, D'Elia,
  de~Ugarte~Postigo, Evans, Even, Fairhurst, Jaimes, Fontes, Fujii, Fynbo,
  Gompertz, Greiner, Hodosan, Irwin, Jakobsson, Jorgensen, Kann, Lyman,
  Malesani, McMahon, Melandri, O'Brien, Osborne, Palazzi, Perley, Pian,
  Piranomonte, Rabus, Rol, Rowlinson, Schulze, Sutton, Thoene, Ulaczyk, Watson,
  Wiersema, \& Wijers}]{Tanvir2017}
Tanvir, N.~R., Levan, A.~J., Gonzalez-Fernandez, C., {et~al.} 2017, ArXiv
  e-prints, 1710.05455

\bibitem[{Wanajo {et~al.}(2014)Wanajo, Sekiguchi, Nishimura, Kiuchi, Kyutoku,
  \& Shibata}]{Wanajo2014}
Wanajo, S., Sekiguchi, Y., Nishimura, N., {et~al.} 2014, \apjl, 789, L39

\end{thebibliography}

%% This command is needed to show the entire author+affilation list when
%% the collaboration and author truncation commands are used.  It has to
%% go at the end of the manuscript.
%\allauthors

%% Include this line if you are using the \added, \replaced, \deleted
%% commands to see a summary list of all changes at the end of the article.
%\listofchanges

\end{document}